\newcounter{myctr}
\def\myitem{\refstepcounter{myctr}\bibfont\noindent\ifnum\themyctr>9\else\phantom{0}\fi\hangindent17pt\themyctr.\enskip}
\def\myhead#1{\vskip10pt\noindent{\bibfont #1:}\vskip4pt}

\documentclass{ws-ijqi}
\usepackage{hyperref}

\usepackage{amsmath}
\usepackage{amssymb}
\usepackage{xcolor}
\usepackage{bbold}
\usepackage[super,sort,compress]{cite}   
\usepackage{ulem}
\usepackage{ulem}

\newcommand{\beq}{\begin{equation}}
\newcommand{\eeq}{\end{equation}}

\newcommand{\bra}[1]{\langle#1|}

\newcommand{\ket}[1]{|#1\rangle}

\newcommand{\id}{\leavevmode\hbox{\small1\normalsize\kern-.33em1}}
\newcommand{\ident}{\mathbb{1}}

\newcommand{\tr}{\textnormal{Tr}}

\begin{document}

\catchline{}{}{}{}{}

\title{
QUANTUM PROCESS MATRICES AS IMAGES: NEW TOOLS TO DESIGN NOVEL DENOISING METHODS
}


\author{MASSIMILIANO GUARNERI}

\address{ENEA - Centro Ricerche Frascati, via E. Fermi 45\\Frascati, 00044, Italy\\
massimiliano.guarneri@enea.it}

\author{ANDREA CHIURI}


\address{ENEA - Centro Ricerche Frascati, via E. Fermi 45\\Frascati, 00044, Italy\\
andrea.chiuri@enea.it}

\maketitle


\begin{abstract}
Inferring a process matrix characterizing a quantum channel from experimental measurements is a key  
issue of quantum information. Sometimes the noise affecting the measured counts brings to matrices very different from the expected ones and the mainly used estimation procedure, i.e. the maximum likelihood
estimation (MLE), is also characterized by several drawbacks. To lower the noise could be necessary to increase the experimental resources, e.g. time for each measurement. In this paper, an alternative procedure, based on suitable Neural Networks, has been implemented and optimized to obtain a denoised process matrix and this approach has been tested with a specific quantum channel, i.e. a Control Phase. This promising method relies on the analogy that can be established between the elements of a process matrix and the pixels of an image.  
\end{abstract}



\markboth{M. Guarneri, A. Chiuri}
{Quantum Process Matrices as Images: New Tools to Design Novel Denoising Methods}

\section{Introduction}	

The characterisation of real-life quantum devices can not entirely relies on the same resources available in a research laboratory, and yet it ought to be reliable and accurate. The operations of these devices rely on the exploitation of quantum properties that should be easily achievable and extractable. Nevertheless, the possibility to fully characterize the quantum features of a system represents still a difficult task that requires a very large number of resources and significant control to provide reliable information. This is particularly true when the complexity of the system is large.

In this framework, also the reconstruction of quantum states, processes and measurement is routinely used to test the performance of quantum devices. Usual direct methods, notably Quantum State Tomography (QST)\cite{jame01pra} and Quantum Process Tomography (QPT) \cite{mohs08pra}, demand a large number of resources to infer the associated matrix enabling the state reconstruction of a quantum system/process when an ensemble of copies identically prepared is available. Indeed, the collected sample should be sufficiently large to provide reliable estimates of the detection probabilities $b_{i,k}=\tr[\Pi_k\rho_i]$ due to the Born's rule. These represent the starting point to infer the state $\rho_i$ from a quorum of measurements $\Pi_k$, and similarly for process and measurement reconstruction.

In some cases, fluctuations and the experimental noise on the measured counts may steer the reconstruction away from the actual matrix, although this effect is in part curtailed by maxlik methods to avoid at least unphysical outcomes.

Another issue is the exponential scaling of quantum tomography that was appreciated since its inception, and has stimulated intense efforts to find ways around this bottleneck [See e.g. Ref.~\citen{bour04prl,ring18prx}]. 

It could be possible to bypass the bottleneck due to the large number of necessary resources and uncertainty of the reconstructed process/state with the development and deployment of the tools used in data intensive research. Specifically, artificial intelligence (AI) could be used to enhance the quality of the acquired data and reduce the number of replica of the experiments, necessary to achieve the desired result, e.g. a parameter extraction.

In this work we present a comparative study between usual maximum-likelihood method and a novel approach based on suitably designed Neural Networks (NNs). 

In literature we can find that these have been employed for quantum tomography, as well as for related parameter reconstruction paving the way to further exploration to understand in full the capabilities and opportunities offered by progress in machine learning \cite{jaeg21nce,muja21} (ML). 
The latter was also applied in several fields of the quantum domain in the last years including state and unitary tomography \cite{spag17sr,palmieri20npj,rocc19sa,arra19qst,torlai18nat,gior20prl,torlai19prl,tiun20opt}, design of quantum experiments \cite{nich19qst,meln18pnas,kren16prl,dris19qmi,saba19pra,gao20prl}, validation of quantum technology \cite{agre19prx,flam19qst,knot16njp}, identification of quantum features \cite{gebh20prr}, adaptive control of quantum devices \cite{hent10prl,hent11prl,love13prl,bona16nat,pali17neuro,liu17pra,paes17prl,lumi18pra,pali19pra,dina19prb,liu20mlst,peng20pra,ramb20prr}, the calibration of a quantum sensor \cite{cimi19prl,cimini21pra}, and the quantum-enhanced pattern recognition \cite{ortolano,pereira}.

Here we demonstrate that the ML method offers relevant opportunities and allow to obtain optimal process matrices without the noise that unavoidably affect those reconstructed from experimental measurements paving the way also to more efficient computational tools. The proposed approach is based on the employment of an Autoencoder Neural Network (aNN), the use of convolutional operations and adaptive filters, and the analogy between the elements of a process matrix and the pixels of an image. We study the optimal configuration of the NNs architecture applying this approach to a Control-Phase quantum channel and we demonstrate that it allows to minimize the requested resources to extract also the parameter characterizing the process.

Employing the method proposed in [\citen{guarneri}], we show that the aNN can play a crucial role also to extract the parameter $\phi$ characterizing the channel with a minimal number of experimental resources and this is true for many values of $\phi$, i.e. for many different processes. The present paper is organized as follows: Section \ref{Sec:materials} consists of an introduction of the different approaches compared in the present manuscript, the description of the test-bench that we employed, i.e. the C-Phase and the simulated noisy processes. In this section some similarities between quantum process matrices and images are demonstrated and it is shown that the correlations between the elements of the matrix play a crucial role to design the desired architecture. This corroborate the use of Convolutional Neural Networks to reconstruct the process matrices characterizing a specific noisy quantum channel. 
Section \ref{Sec:methods} discusses in detail the denoising procedures that we applied to the noisy processes with a specific focus on the aNN. The results obtained with the two methods and their comparison are reported in Section \ref{Sec:result}, while in Section \ref{Sec:parm_extr} we show the capabilities of the proposed ML approach for quantum parameters extraction. The conclusions are given in Section \ref{conc}.


\section{Noisy processes generation and possible denoising procedures}

\label{Sec:materials}
\subsection{Dataset generation and employment} 
\label{Sec:dataset_gen}
\textit{Quantum Process Tomography (QPT) - } A quantum process acting on a system described by a density matrix $\rho$ associated to a $d$-dimensional Hilbert space can be fully characterized by a Kraus representation \cite{mohs08pra,Chuang97,nielsen,qpt-exp} as follows:
\begin{equation}
\varepsilon(\rho) = \sum_i E_i \rho E_i^{\dagger}
\end{equation}
where $E_i$ are operators acting on the system and $\sum_i E_i E_i^{\dagger} = \ident$; hence, the quantum process tomography of $\varepsilon$ consists of the experimental reconstruction of the operators $\{E_i\}$. Each operator $E_i$ can be related with measurable parameters using a basis of operators $\{A_i\}$, i.e. $E_i = \sum_m a_{im} A_m$. It follows that the process can be written as follows:
\beq
\varepsilon (\rho_{in}) = {\sum}_{m,n} \chi_{mn} A_m  \rho_{in} A^{\dagger}_n
\eeq
where $\chi_{mn} = \sum_i a_{im} a^{*}_{in}$. To infer the matrix $\chi_{mn}$ from experimental data, a basis for the Hilbert space of $d \times d$ matrices should be prepared, i.e. $d^2$ input states $\rho_k$. The output states read as   
\beq
\varepsilon(\rho_k) = {\sum}_{j} \lambda_{kj} \rho_{j}
\eeq

Here the coefficients $\lambda_{kj}$ can be experimentally obtained exploiting the basis $\{\rho_k\}$ previously defined. Indeed, introducing the coefficients $\beta^{mn}_{jk}$ as follows $A_m  \rho_{j} A^{\dagger}_n=\sum_k \beta^{mn}_{jk} \rho_k$, the coefficients $\lambda_{kj}$ can be related to the matrix elements $\chi_{mn}$:
\beq
\sum_{mn} \beta^{mn}_{jk} \chi_{mn} =\lambda_{jk}
\eeq
Inverting the matrix $\beta^{mn}_{jk}$, it is possible to find the process matrix $\chi$:
\beq
\chi_{mn} =\sum_{jk} \tau^{mn}_{jk} \lambda_{jk}
\eeq
where $\tau^{mn}_{jk}$ is the inverse matrix, i.e. $\tau^{pq}_{jk} \beta^{mn}_{jk} = \delta_{pm} \delta_{qn} $.

\textit{The Control Phase (CP) - } A CP is a 2-qubit controlled quantum gate characterized by a single parameter, i.e. $\phi$, defined as follows:

\begin{equation}
\label{Tmatrix}
CP(\phi)= \left(
\begin{array}{cccc}
1& 0  &  0  &  0\\

0  & 1 &  0  & 0 \\

0  &  0  &  1 & 0 \\

0  &  0  &  0  &  e^{- i \phi}
\end{array}
\right).
\end{equation}
and ideal $CP(\phi=\pi)\equiv CZ$ gate can be decomposed as a coherent sum $CP(\phi=\pi) = (\id \otimes \id + \id \otimes \sigma_Z +  \sigma_Z \otimes \id - \sigma_Z \otimes \sigma_Z)/2$ of tensor products of Pauli operators $\{\id, \sigma_X, \sigma_Y, \sigma_Z\}$ acting on control and target qubits respectively. The process $\chi$ related to a general $CP(\phi)$ gate represents an interesting test-bench for the purposes of this manuscript because it is a 16x16 matrix characterized by 256 elements \cite{white07josab} comparable with the  pixels of a low-resolution image. Here we do not show all the possible process matrices $\chi$ because they depend on the value of $\phi$, we only provide few examples.

\begin{figure}[h]
\centerline{\includegraphics[width=\columnwidth]{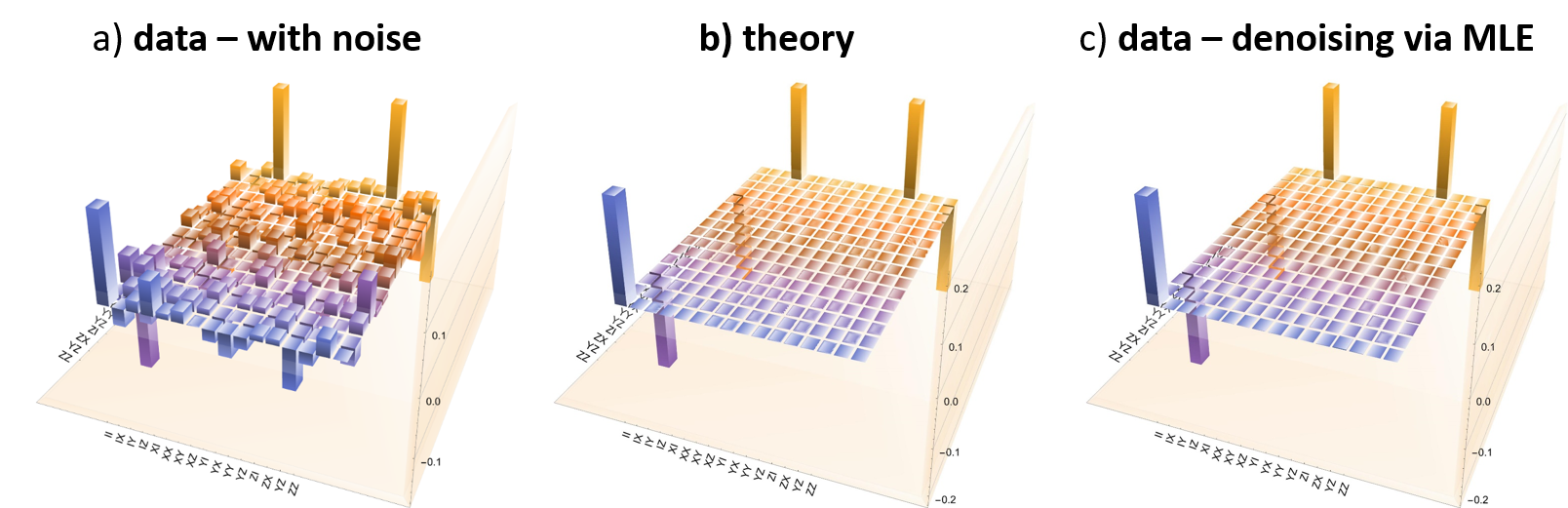}}
\vspace*{8pt}
\caption{a) b) expected matrix, c) denoised matrices obtained implementing a MLE algorithm.}
\label{fig:results}
\end{figure}

\textit{The noisy matrices - } We have generated a dataset composed by 500 processes matrices for each value of $\phi$ reconstructing the associated $\chi$ matrix \cite{white07josab,qpt-exp,cresp11natcom}: 
\begin{itemize}
\item We created a suitable basis of two qubit states given by the tensor products of single
qubit states. 16 different input states were considered and the projections of the output states were calculated on the set $\{ \ket{i,j} \}$ where $i,j = \ket{0}, \ket{1}, 1/\sqrt{2}(\ket{0}\pm \ket{1}), 1/\sqrt{2}(\ket{0}\pm i \ket{1})$  and $\{\ket{0}, \ket{1}\}$ is the orthonormal basis used to represent the state of a qubit.

\item we have considered 16 different values of $\phi= \pi/6$, $ \pi/4$, $ \pi/3$, $ \pi/2$, $ 2\pi/3$, $ 3\pi/4$, $ 5\pi/6$, $ \pi$, $ 7\pi/6$, $ 5 \pi/4$, $ 4\pi/3$, $ 3\pi/2$, $ 5\pi/3$, $ 7 \pi/4$, $ 11\pi/6$, i.e. 16 different processes, and 500 instances for each value.
\item we evaluated the corresponding expected counts for each measurement and we added a Poissonian noise to each value
\item we repeated this procedure for three different level of signal (i.e. N=2000, 1000, 200 counts/s as total resources; $N=r_i \cdot 2000$, $r_i=1,0.5,0.1$ for i=1,2,3). 
\item we have estimated the experimental matrix which can be obtained directly by the simulated counts via the QPT [See an example in Fig. \ref{fig:results}a)] and these processes composed the considered \text{noisy} dataset. 
\end{itemize}

\textit{Dataset employment - } This dataset was used to apply and test a usual MLE algorithm [See an example of the obtained matrices in Fig. \ref{fig:results}c)] and a novel optimized denoising method \cite{guarneri} based on an Autoencoder Neural Network (aNN) [More details in Section \ref{aNN:intro1},\ref{aNN:intro2} and \ref{Sec:aNN}]. Precisely, the simulated matrices and the corresponding theoretical ones were used to train and test the aNN.

\subsection{Quantum matrices as two layers images}
\label{aNN:intro1}

\begin{figure}[h]
\centerline{\includegraphics[width=\columnwidth]{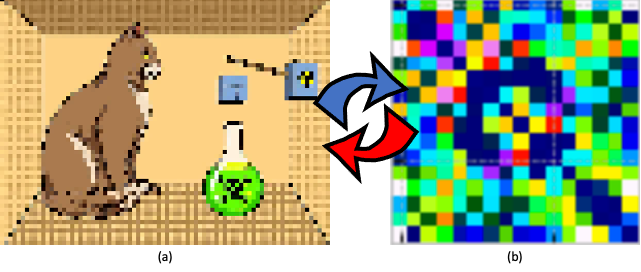}}
\vspace*{8pt}
\caption{Analogy between coloured images (a) and a CPhase process matrix (b).}
\label{analogia}
\end{figure}

In computer graphics, an image is a combination of different layers (usually three: red green and blue), each one composed by a matrix of NxM pixels, where each value represents the intensity of the considered channel. A quantum process is represented by two layers (real and imagery parts of a complex matrix), each one composed by a NxN matrix. A real quantum process is usually affected by noise, which can compromise the desired outcome of the experiment or the evaluation of some parameters characterizing the channel. An intriguing possibility is represented by the use of image processing algorithms to obtain denoised process matrices. In particular, we tried to use the analogy between images and quantum complex matrices [See Fig. \ref{analogia}] for investigating the potentiality of the Convolutional Neural Networks (CNN) as a valid tool.
If we focus on the effect of the Poissonian noise introduced in the quantum matrices [See Sec.\ref{Sec:dataset_gen} for more details], it is possible to observe some symmetries. This reinforced the idea that the application of image processing algorithms could represent a suitable tools to be applied also in the quantum domain.

\begin{figure}[h]
\centerline{\includegraphics[width=\columnwidth]{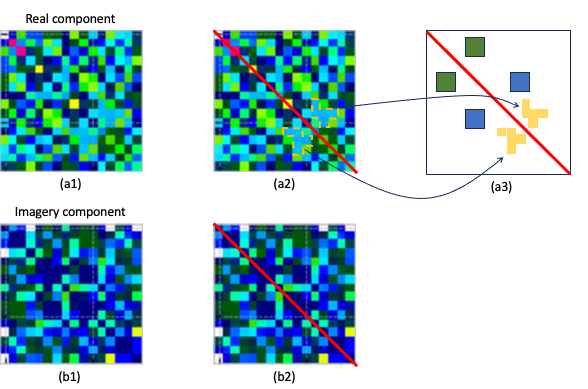}}
\vspace*{8pt}
\caption{An example of symmetry in the resulting matrix obtained by subtracting theoretic and noisy quantum channel matrix. a1 and b1 are respectively the resulting real and imagery components. The red lines on the main diagonal of each matrix show the specular symmetry axis of each element belonging to the upper and lower part (a2 and b2)}
\label{theoric_minus_noisy}
\end{figure}

Figure \ref{theoric_minus_noisy} shows these symmetries obtained as the difference between the noisy matrices and the theoretical ones (a1,b1). The symmetric axis is the main diagonal of each matrix (a2,b2): an example is shown in (a3). These properties suggested that it was possible to consider these matrices as images (in particular as pixel-art images) and to apply typical algorithms employed in this domain, i.e. Neural Network denoising techniques. These can ensure several advantages, e.g. the possibility to minimize the possibility to extract wrong parameters, the reconstruction of denoised processes without a significant increasing of the necessary quantum and computational resources or time. 

\subsection{Autoencoder Neural Networks}
\label{aNN:intro2}
One of the most powerful and diffused Neural Network (NN) is the so called Autoencoder NN. This NN has an architecture composed by a first block, called Encoder, which is able to code the input in a lower dimensional vector and a second block, named Decoder, able to reconstruct the original signal from the coded vector obtained as output from the Encoder. If we limit the use of the autoencoder just to this operation, the result will not be better than any other compression algorithm. On the other side, the particularity of this network to reduce the dimensionality of the input to a vector of features can be used to reduce the noise of the original signal. 

\begin{figure}[t]
\centerline{\includegraphics[width=\columnwidth]{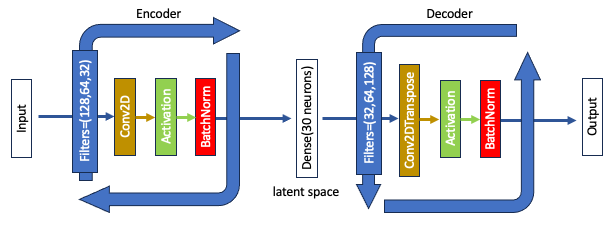}}
\vspace*{8pt}
\caption{aNN scheme: the Encoder is composed by three layers, looping on three different filters; the decoder has the same structure, but the loop is made on the opposite sequence of filters and convolution layer is substituted by its homologous transposed; the latent space, where the features of the status matrices are extracted, connects the Encoder and Decoder architectures. }
\label{autoencoder_schema}
\end{figure}

This kind of network is mainly diffused in image processing field for denoising, restoration and even colorization of old pictures\cite{Autoencoders}. 
Starting from a standard architecture for the aNN, where the Encoder is composed of three convolutional layer and the Decoder of an equivalent number of transposed convolutional layer, we chose the hyperparameters by an empirical process identifying the best filters' kernel size and features vector dimensionality as described in Sec.\ref{Sec:aNN}. Precisely, a range of kernel values was considered to understand the role of the correlations of the matrix elements within the architecture of the employed NN and to define the best configuration allowing to perform an optimal denoising.

A scheme of the aNN architecture used in this work is reported in Fig. \ref{autoencoder_schema}. 

\subsection{Maximum Likelihood Estimation (MLE)}
\label{Sec:MLE}

The MLE method aims at finding the matrix that maximize the probability of the observed experimental
data, by restricting the search to semidefinite positive and unit trace density matrices. Therefore,
to get the MLE, a constrained optimization problem has to be solved. 
The optimization difficulty increases with the dimension of the matrix to be reconstructed. 
Alternative approaches leading to full optimization have been proposed in the case of 
some prior knowledge of the state. 
The constraint problem may be solved by parameterizing the matrix in order to ensure both
positivity and unit trace, and by solving the new unconstrained optimization \cite{jame01pra} as explained in this Section. 

\textit{The MLE method - } The measurement data consist of a set of $\nu$ coincidence counts $n_i$ ($n_i$=1,2, . . . ,$\nu$) whose expected value is $\overline{n_i}= N \bra{\psi_{i}} \hat{\rho} \ket{\psi_i}$. 
Assuming a Gaussian probability distribution for the coincidence measurements, the probability of obtaining a set of $\nu$ counts
$\{n_1 ,n_2 , . . . n_{\nu} \}$ is expressed as
\noindent
\begin{equation}\label{likel}
 L(\hat{\rho})= p(data| \hat{\rho}) = \frac{1}{N} \prod^{\nu}_{i=1} exp[-\frac{(n_i-\overline{n_i})^2}{2 \sigma_i^2}]
\end{equation}

where $\sigma_i$ is the standard deviation of the $n_{th}$ coincidence
measurement characterized by a Poissonian distribution (i.e. $\sqrt{n_{th}}$) 
and $N$ is the
normalization constant.
The main goal of the MLE is to  find the maximum of this probability and this corresponds to maximize the probability of getting the observed data for a 
given quantum state $\hat{\rho}$.

In order to satisfy the necessary mathematical properties of a physical matrix, which should be 
Hermitian semidefinite positive and with unit trace,
the density  matrix must be parametrized as: 
\noindent
\begin{equation}
\hat{\rho}(t)= \frac{T^{\dagger}(t)T(t)}{Tr[T^{\dagger}(t)T(t)]}
\end{equation}
with 
\begin{equation}
\label{Tmatrix}
T(t)= \left(
\begin{array}{ccccc}
t_1 & t_{d+1} + i t_{d+2}  &  ...  &  ... & t_{d^2-1} + i t_{d^2}\\

0  & t_2 &  t_{d+3} + i t_{d+4}  & ... & ...\\

0  &  0  &  ... & ...  & ..\\

...  &  ...  &  ...  & ...& t_{3d-1} + i t_{3d-2}\\

0  &  ...  &  ...  &  0 & t_d
\end{array}
\right).
\end{equation}
where $d$ is the dimension of the state space.\\



The MLE method consists of the maximization of $L(\hat{\rho}(t))= p(data| \hat{\rho}(t))$, i.e. of
finding the set of parameters $\{t^*_1,...,t^*_{\nu}\}$ which maximizes the function $L(n_{1},..,n_{\nu} | t_{1},..,t_{\nu})$. 
The density matrix then will be $\hat{\rho}(t^*_1,...,t^*_{\nu})$.

Generally people used to find the minimum of $Log[L(\hat{\rho}(t))]$, namely the {\it log-likelihood function} instead of 
the maximum of $L(\hat{\rho}(t))$ given by a product of exponential functions [See Eq.(\ref{likel})].

\textit{MLE limits - } 
The optimization problem previously described has a non convex objective function with
several local minima. The equivalence among local solutions of the associated unconstrained 
optimization problem has been demonstrated in \cite{ribe12qic}.
Hence, the fact that the function $L(\rho(t))$ is not convex does not represent a problem since any minimum is sufficient to reconstruct the density matrix of the unknown quantum state and a local no-global minimum does not lead to a wrong estimation.

In the same paper it was shown that some errors in the reconstruction may be caused by the use of numerical optimization methods where 
global convergence to local minimizers does not hold or stop prematurely,
indicating the stagnation of the optimization process.

It is worth noting that MLE typically yields to one or more
zero eigenvalues of the reconstructed density matrix. Such an estimation can be difficulty accepted  since it implies that some measurement outcomes are completely impossible. No finite amount of data can justify such certainty. 
As a consequence the results achieved by MLE are sometimes unreliable\footnote{This issue becomes relevant with quantum 
states involving two or more qubits.}.

A quantum state is somehow a prediction of future. Similarly to a classical
probability distribution, it predicts probabilities for all the measurements that can be performed. It follows that it is unbelievable that some measurement outcomes
are predicted as impossible. 
Even if MLE allows to make physical the measured density matrices, the 
discussed issue still remains a challenging topic to be faced.

Moreover, by considering a quantum state with one or more zero eigenvalues, it is difficult to give an interpretation of the error on the estimated matrix $\hat{\rho}$. Indeed, in these cases $\hat{\rho}$ lies near the frontier of the physical states space and  $\hat{\rho} \pm \Delta \hat{\rho}$ defines a ensemble of state containing also non physical matrices.  

\section{Methods}
\label{Sec:methods}

\subsection{Denoising via AutoEncoder Neural Network}
\label{Sec:aNN}
The aNN that we employed is composed of three convolutional layers (and equal number of transposed convolutional), with 128, 64 and 32 filters, respectively. 
These layers were introduced to exploit the correlations characterizing the elements of the considered matrices, e.g. among the real and imagery parts. 
Each filter was composed of a kernel size $k_i \times k_i$ where $k_i=1...7$ and the optimal configuration will be discussed at the end of this Section. The activation function, after each convolutional layer, was the Rectified Linear Unit (ReLU): it  is a piecewise linear function that outputs the input directly if it is positive, otherwise it outputs zero. 
For speeding up the training process and avoiding the overfitting of the network, a batch normalization layer was introduced after every activation function of each convolutional (and transposed convolutional) layer\cite{batchnorm}.

The training phase was computed with a dataset of $24000$ complex matrices composed by the simulated noisy processes obtained as described in Sec.\ref{Sec:dataset_gen}, the $75\%$ was used to train the net, the $10\%$ to validate it, and the final $15\%$ was used as final test. For this work, each quantum process matrix was stored in a file as a unidimensional vector, where each term is a complex number: as described above, to maintain the analogy with the image processing domain and facilitate an easiest processing of the complex matrices by the autoencoder net, a reshaping of every vector was needed: $\mathbb{C}^{1x256} = \mathbb{C}^{16x16} = \mathbb{R}^{16x16x2}$.

To allow the use of the ReLU, the dataset was normalized before inputting in the aNN, avoiding the negative values; the resulting output was rescaled at the correct values by using the minimum and maximum values of the dataset. Hence, given the dataset $N=\{X_k|x_{i,k}=a_{i,k}+j\cdot b_{i,k}$, with $i=1\dots 256, k=1 \dots 18000\}$, each element of the matrix was normalized following the steps:
\begin{align}
\label{eq:norm}
\begin{split}
    m=min(N)
    \\
    M = max(N)
    \\
    \hat{x}_{i,j} = \frac{x_{i,j}-m}{M-m}
\end{split}
\end{align}
where $\hat{x}_{i,j}$ represents the normalized element. The output $D=\{Y_k|y_{i,k}=a_{i,k}+j\cdot b_{i,k}$, with $i=1\dots 256, k=1 \dots 2000\}$ obtained by the trained autoencoder, processing a noisy quantum process matrices dataset, was rescaled following the equation:
\begin{align}
\label{eq:rescal}
    \hat{y}_{i,j} = m + y_{i,j} \cdot (M-m)
\end{align}
where $\hat{y}_{i,j}$ represents the $i-th$ element of matrix resulting from the rescaling.

\begin{figure}[t]
\centerline{\includegraphics[width=\columnwidth]{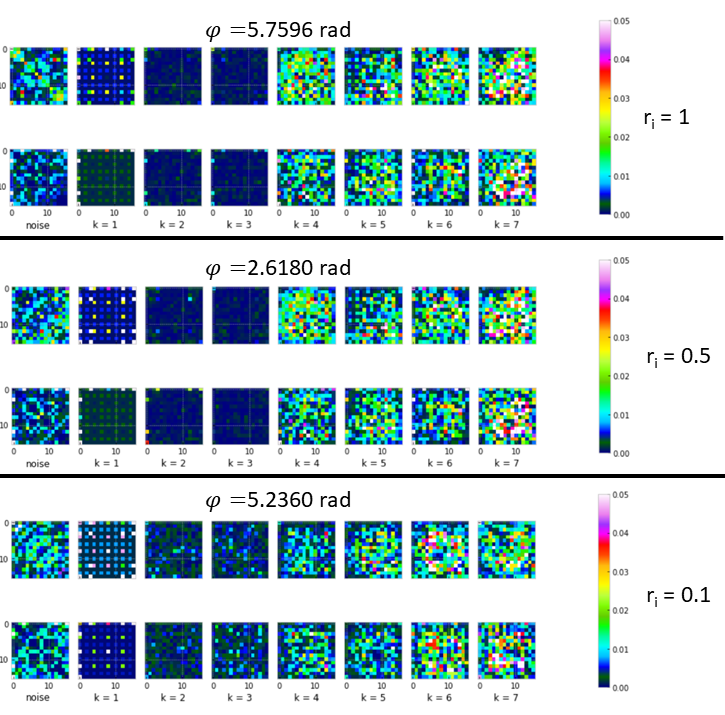}}
\vspace*{8pt}
\caption{Heatmaps of some denoised process matrices for several values of the kernel size. The first column represents the difference between the theoretic and noisy matrices, while each row is the  real and imagery components, respectively. These results where obtained for the three different signal levels, i.e. $r_i=0.1,0.5,1$.}
\label{denoise_vs_kernel}
\end{figure}

After this prepocessing, several autoencoders, obtained by changing the kernel size $k_i$ of the convolutional filters, were trained by these data. This study was aimed at selecting the best kernel-size which minimize the difference between the theoretical matrices and the resulting ones coming from the denoising process realized by the different trained autoencoders. Figure \ref{denoise_vs_kernel} shows some examples of the heatmaps resulting from the difference between the theoretical and denoised matrices, estimated by each aNN described above, i.e. for $k_i=1...7$. The first column shows the differences between the theoretical and noisy matrices without the employment of the aNN. Here it is possible to observe the symmetries previously discussed. Based on these results it was possible to define the best kernel-size for $k_i=2,3$. Thus, the trivial choice $k_i=1$ is not the optimal one and the correlations between the elements of the matrix can be suitably exploited to improve the action of the Convolutional Neural Networks. Although the obtained number cannot be generalized to all the possible quantum channels, the same approach can be certainly extended.

\subsection{Denoising via MLE}

For different values of $\phi$ we reconstructed the experimental $\chi$ matrix of the process as described in Sec.\ref{Sec:dataset_gen} and we performed an optimization of these matrices employing the usual maximum-likelihood approach \cite{obrien04,jezek03} as described in Sec. \ref{Sec:MLE}, i.e. finding the set of parameters $\{t^*_1,...,t^*_{\nu}\}$ which maximizes the function $L(\hat{\rho}(t))= p(data| \hat{\rho}(t)) \rightarrow L(n_{1},..,n_{\nu} | t_{1},..,t_{\nu})$. The resulting density matrices then read as $\hat{\rho}(t^*_1,...,t^*_{\nu})$.

\section{Results: comparison between the two methods}
\label{Sec:result}




\begin{figure} 
	\begin{minipage}{\columnwidth}
	\centering
	 \includegraphics[width=\columnwidth]{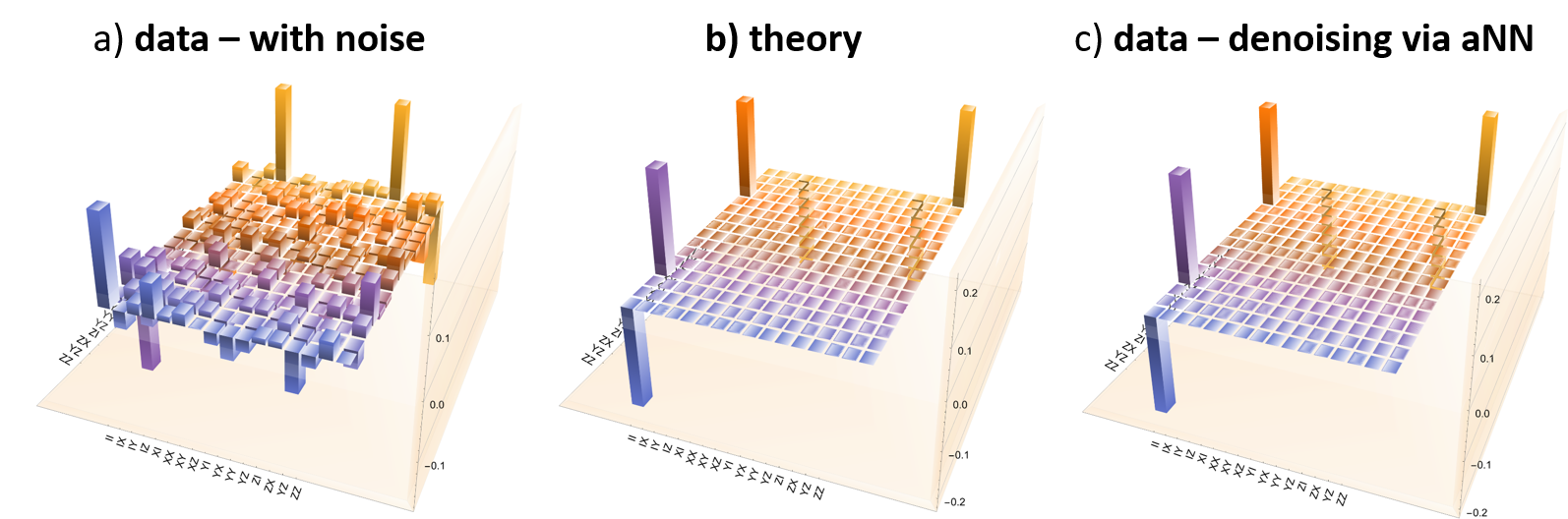}
	 \end{minipage}
	 	\vspace{1em}
	 \caption{$Im[\chi_{CP}$($\phi= 5\pi /3$)]. a) matrix obtained directly by the simulated counts via the QPT; b) theoretical process matrix; c) matrix obtained as output of the aNN.}  
\label{resultsCPHASE1} 
	\vspace{1em}
\begin{minipage}{\columnwidth}
\centering
	 \includegraphics[width=\columnwidth]{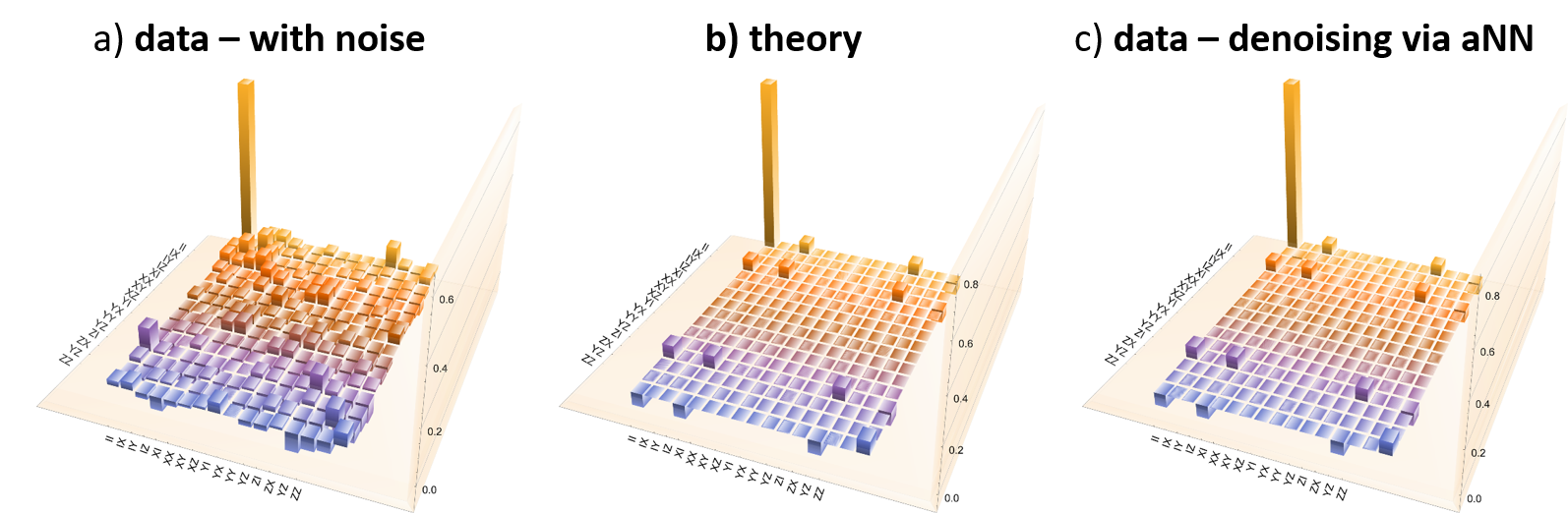}
	 \end{minipage}
	 	\vspace{1em}
	  \caption{$Re[\chi_{CP}(\phi= 5\pi /3)] $. a) matrix obtained directly by the simulated counts via the QPT; b) theoretical process matrix; c) matrix obtained as output of the aNN.} 
\label{resultsCPHASE2}
	\vspace{1em}
	\begin{minipage}{\columnwidth}
	 \centering 
	 \includegraphics[width=\columnwidth]{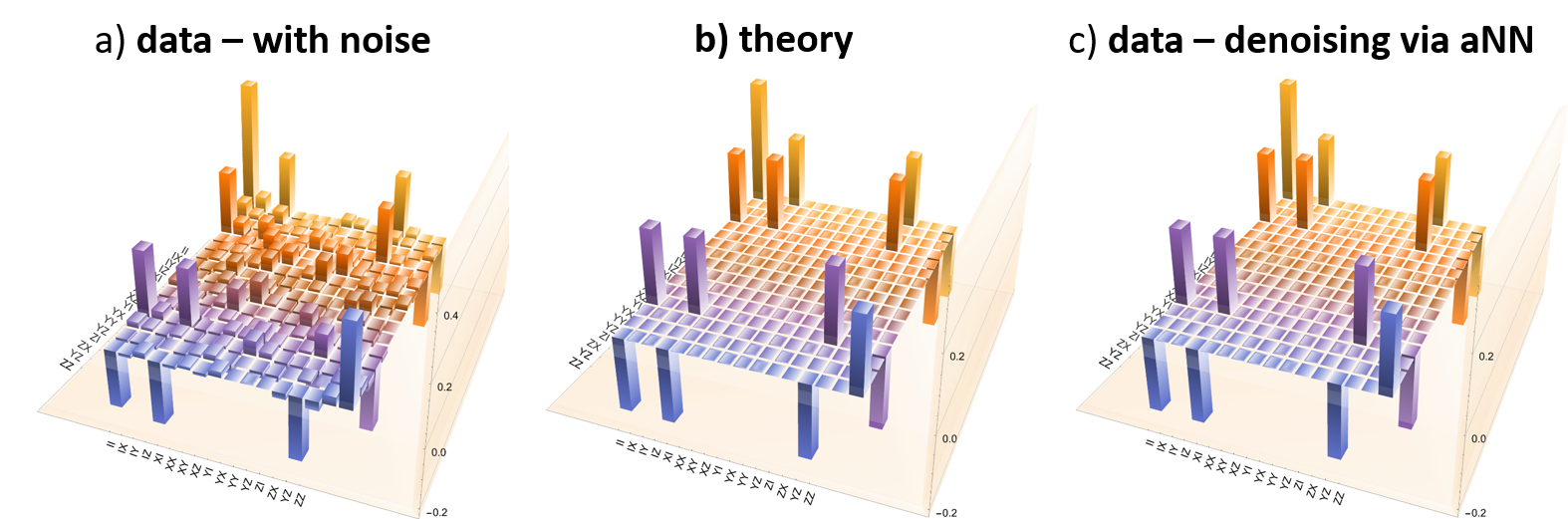}
	\end{minipage}  
	\vspace{1em}
	  \caption{$Re[\chi_{CP}(\phi= 3\pi /4)]$. a) matrix obtained directly by the simulated counts via the QPT; b) theoretical process matrix; c) matrix obtained as output of the aNN.} 
    \label{resultsCPHASE3}
    	\vspace{1em}
\end{figure}

{\it Process Matrices.} In Fig. \ref{resultsCPHASE1}, \ref{resultsCPHASE2}, \ref{resultsCPHASE3} we show the matrices obtained by employing the aNN for several simulated noisy CP channels, i.e. different values of $\phi$. For each process we also report the expected theoretical matrix.

{\it Fidelity.} In order to evaluate the quality of the obtained data we have calculated the process fidelity \cite{cresp11natcom} as:

\begin{equation}
F_{exp} \propto Tr[\sqrt{\sqrt{\chi_{exp}} \chi_{CPhase} \sqrt{\chi_{exp}} }]^2
\label{this}
\end{equation}

where $\chi_{CPhase}$ is the process matrix of the ideal CPhase gate.

We compared the fidelities resulting from the application of the aNN and  MLE algorithms, respectively. Both of them perform a denoising over the noisy data but employing the aNN it is possible to achieve better performances: the Fidelity \cite{jos94jmo} with the expected matrix is always $\geq 0.99$ with large amount of values  $= 1$ and this is true also when we deal with very noisy data, i.e. $r_3=0.1$. With the usual MLE algorithms we obtained worse results, i.e. Fidelity generally $< 0.99$ [See Table.\ref{Table_results}].

\begin{table}
\caption{Results obtained for some values of $\phi$. We have tested our approach over a sample of $\approx 30$ 
noisy matrices and we report here the mean value of the Fidelity for the two considered methods. i.e. aNN and MLE. The table shows the results for $r_3=1$. }
\centering
\begin{tabular}{c | c c }
\hline
\hline
$\phi$ & $F_{aNN}$ & $F_{MLE}$\\
\hline
$5\pi/4$  & $1.000 \pm 0.002$  &  $0.991 \pm 0.003$ \\
$5\pi/3$ & $1.000 \pm 0.001$  &  $0.991 \pm 0.002$\\
$5\pi/6$  & $0.992 \pm 0.002$  &  $0.991 \pm 0.003$\\
$\pi/2$  & $1.000 \pm 0.002$  &  $0.981 \pm 0.007$\\
$\pi/6$  & $1.000 \pm 0.001$  &  $0.991 \pm 0.003$\\
\hline
\hline
\end{tabular}
\label{Table_results}
\end{table}

\section{Machine Learning for parameter extraction}
\label{Sec:parm_extr}

The denoised matrices, i.e. the output of the aNN, were used as inputs of a dense Feed-Forward NN (FFNN) to extract the parameter characterizing the considered channel as described in \cite{guarneri}. The adopted strategy relies on the introduction of a denoiser stage,i.e. the aNN, aimed at denoising the input matrices, and then a dense multilayers network (i.e. the FFNN) allows to extract the parameter.
The FFNN was chosen with a modular structure: the last layers of this network are forked with the same number of parameters to estimate for single process or status matrix. This modular architecture allowed to improve the parameters estimation if compared with a network characterized by common dense layers and a single multi-neural output layer.

\begin{figure*}[h]
\includegraphics[width=\textwidth]{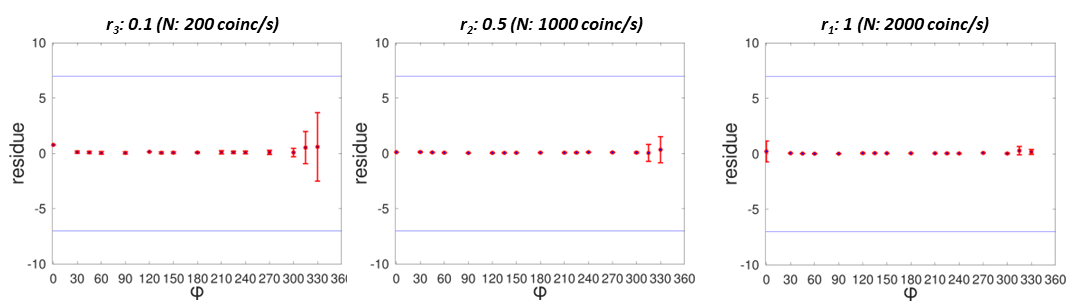}
\caption{Residue of the extracted value of $\phi$ compared to the expected one. The proposed approach achieves a success rate $\approx 100 \%$ even for the lowest signal level, $r_3=0.1$. If the obtained residues belong to the range defined by the blue lines, i.e. [-7,7], it is possible to discriminate among the 16 considered values of $\phi$.}
\label{results_param}
\end{figure*}

Here we show new results, summarised in Fig.~\ref{results_param}, where it is confirmed that, even for the lowest signal level at $r_3=0.1$ it is possible to discriminate among the 16 different values of the parameter $\phi$. 

The quantum parameters extraction was made by pre-training a Neural Network, where in input theoretical process and status matrices and in output the expected parameters were passed. As described in the previous Sections, noisy matrices were obtained by introducing a Poisson noise into simulated processes and their related counts. 

During the training phase, the two NNs composing the employed architecture were trained separately: this choice allowed to obtained a more robust control of the entire training and fine-tuning processes during the denoising and estimation phases.

\section{Conclusions}
\label{conc}

The results presented in this paper demonstrates that the capabilities offered by the aNNs in data denoising can be fully exploited also in a completely different scenario, i.e. the reconstruction of noisy process matrices characterizing a quantum channel. In this paper we have bench-marked this approach and compared the performance to those achievable with usual MLE algorithms. We have demonstrated that the application of the proposed architecture to a real quantum process can guarantee a benefit over the usual methods. The employed architecture relies on the analogies between a quantum matrix and an image, but also on the correlations between the elements of the matrix. Indeed the latter was characterized and the kernel size of the NN was studied to understand its role in the designed architecture. We showed that a stable and reliable denoising is achievable even when a low number of resources is considered and by training the network with only simulated data. The output of the aNN was also used as input of suitably designed FFNN and the parameter $\phi$ was extracted with high accuracy. Here we showed some interesting results that allow to demonstrate the feasibility of this method in non-trivial scenarios, when the parameter characterizing the channel is not related to the element of the process matrix. The obtained results show the viability of this approach as an effective tool based on a completely new paradigm for the employment of NNs in the quantum domain.

\section*{Acknowledgments}
This work was supported by the NATO SPS Project HADES - MYP G5839.


\end{document}